\documentclass[a4paper]{article}
\usepackage{amsmath,amssymb,mathrsfs}
\usepackage{gfsartemisia}
\usepackage[T1]{fontenc}
\usepackage[top=3cm,bottom=3cm,left=3cm,right=3cm]{geometry}
\date{}
\begin{document}
\author{Burak Tevfik Kaynak${^\dag}$ and O. Teoman Turgut$^\ddag$ \\ Department of Physics, Bo\u{g}azi\c{c}i University \\ 34342 Bebek, Istanbul, Turkey \\ $^\dag$burak.kaynak@boun.edu.tr, $^\ddag$turgutte@boun.edu.tr \\
\small \emph{Dedicated to the memory of Professor Yavuz Nutku}}
\title{\bf Singular interactions supported by embedded curves}
\maketitle
\begin{abstract}
In this work, singular interactions supported by embedded curves on Riemannian manifolds are discussed from a more direct and physical perspective, via the heat kernel approach. We show that the renormalized problem is well defined, the ground state is finite and the corresponding wavefunction is positive. The renormalization group invariance of the model is also discussed.
\end{abstract}
\section{Introduction}
Schr\"{o}dinger operators with singular interactions have been studied in depth for some time by now. The literature on this subject is vast, and the reader is invited to refer to~\cite{al1,al2} and references therein for extensive studies on this subject. On the other hand, similar problems in which interactions are supported by curves attracted the attention of mathematicians and mathematical physicists as a result of~\cite{bra,ex1}. The physical motivation behind these studies stems from the need for modeling semiconductor quantum wires~\cite{ex2}, in which the Schr\"{o}dinger operator with singular interactions supported by curves were introduced as a model for quantum mechanics of electrons confined to narrow tube-like regions. Following these works, the theory of curve-supported singular interactions in $\mathbb{R}^3$ from the point of view of self-adjoint extensions was developed rigorously in~\cite{ex3,ex4,ex5,ex6,ex7,lob} and references therein. These works also investigated the positivity of the ground-state wavefunction and the regularity of the eigenfunctions. The case of periodic curves was also thoroughly investigated from the same perspective. In general, the problem admits an infinite number of extension parameters, but it is natural to choose a single-extension parameter for simplicity, as we have also done, as a result, the value of the emerging coupling constant is used to define the theory. Therefore, it is natural to investigate the range of parameters, which would lead to bound states. In this work, we follow a more direct approach, which is more physical and less rigorous, and we choose the bound-state energy of the curve as the defining parameter.   

The aim of this work is to study the Schr\"{o}dinger operator with singular interactions supported by curves embedded in a Riemannian manifold. We consider finite-length closed curves, and they are neither intersecting nor self-intersecting. The curves under consideration do not have any dynamical degrees of freedom. Moreover, we use only one generic coupling constant in order to define the strength of the interactions. Therefore, we will analyze how a quantum-mechanical particle interacts with singular interactions supported by stationary curves. Our primary motivation in studying this model is to test various nonperturbative renormalization methods so as to extend our understanding of the renormalization program for the models in Riemannian manifolds. As the bound-state problem is nonperturbative, we construct the resolvent for the full Hamiltonian in terms of the so-called principal operator as in~\cite{raj,et}. The distinctive advantage of this formulation is that once the renormalization of this operator is accomplished, we can easily scrutinize various crucial aspects of the model just by working out that operator, such as whether the ground-state energy is bounded from below or not, the positivity of the ground state and the renormalization group equation, which the principal operator is expected to obey. The main result of this work is that the renormalization of the model can be performed nonperturbatively. We will see that one needs to choose a certain prescription in order to renormalize this operator for the bound-state problem, whereas it is more appropriate to utilize another prescription in order to study the renormalization group equation. If we think of renormalization as a way of writing down a sensible effective theory, this study shows that this can be performed successfully, i.e. keeping all the desired features of quantum mechanics intact. The details of the small scale physics are encoded, in some sense, by a minimal modification of the Hamiltonian.

The organization of the paper is as follows. In section~\ref{s2}, the model will be constructed. The renormalization regarding the bound state structure will be studied. In section~\ref{s3}, the fact that the ground state is bounded from below will be shown by using the Ger\u{s}gorin theorem. In section~\ref{s4}, the positivity of the ground-state will be proven, which resides on the Perron-Frobenius theorem. In section~\ref{s5}, the renormalization group equation will be obtained, and it will be demonstrated that the model is asymptotically free as its beta function is found to be negative. In section~\ref{s6}, the two-dimensional case will briefly be discussed.            
\section{Construction and renormalization of the model} \label{s2}
In this section, the renormalization of the model will be studied. The model in which we are interested is given by a generalized Schr\"{o}dinger operator with a singular interaction, whose support is an arclength-parametrized closed curve $\gamma(s)$ of length $L$ embedded in a three-dimensional Riemannian manifold. Its Schr\"{o}dinger equation can formally be written as 
\begin{align}
-\frac{\hbar}{2 m} \nabla_g^2 \psi(x) - \frac{\lambda}{L} \int_\Gamma d_g s \, \delta_g \left( x , \gamma(s) \right)  \int_\Gamma d_g s' \, \psi \left(\gamma(s') \right) &= E \psi(x) \,,
\end{align}
where $\Gamma$ stands for the curves over which the integrals are taken, and $\lambda$ is the coupling constant. In general, we may have various non-intersecting curves, which we label as $\Gamma_i$, and their couplings may labeled as $\lambda_i$, as well. We assume that each $\Gamma_i$ has a finite length, and they are not self-intersecting. Moreover, they have to stay a certain minimum distance away from each other, which we may call $d_{ij}$, and they are only allowed to come close to themselves up to a certain distance, let us say $\Delta$, as we will make more precise later on.

Following~\cite{et}, we need to obtain the regularized resolvent so that the principal operator can be used to renormalize the theory. This could be made most efficiently as follows. We introduce a family of functions supported on curves: 
\begin{equation}
\Gamma_i^\epsilon(x)=\int_{\Gamma_i} d_g s K_{\epsilon/2}(x, \gamma_i(s)) \,.
\end{equation}
Note that as $\epsilon\to 0^+$ we get a delta function supported on the curve. Also,
\begin{equation}
   \langle \Gamma_i^\epsilon\vert \Gamma_i^\epsilon\rangle=\int_{\Gamma_i\times \Gamma_i} d_g s \, d_g s' K_{\epsilon}(\gamma_i(s), \gamma_i(s')) \,,
\end{equation}
which is finite since $K_\epsilon$ is finite at all points as long as $\epsilon>0$. We can rewrite a regularized Schr\"{o}dinger equation for this family,
\begin{align}
(H_0 - E) \vert \psi \rangle = \sum_i \frac{\lambda_i}{L_i} \vert \Gamma_i^\epsilon \rangle \langle \Gamma_i^\epsilon \vert \psi \rangle \,,
\end{align}
in which the integrals over curves are suppressed, and the interaction is written as a projection since we restrict ourselves only to the case of non-intersecting curves. In order to obtain the resolvent, one needs to solve for $\langle \tilde{\Gamma}_i^\epsilon \vert \psi \rangle$ in the following expression for the resolvent:
\begin{align} \label{psi}
\vert \psi \rangle &= (H_0 - E)^{-1} \vert \tilde{\Gamma}_i^\epsilon \rangle \langle \tilde{\Gamma}_i^\epsilon \vert \psi \rangle + (H_0 - E)^{-1} \vert \varphi \rangle \,, 
\end{align}
in which $\sqrt{\lambda_i /L_i}$ is absorbed in $\Gamma_i^\epsilon$, and a summation is assumed over repeated indices. It is easy to show that
\begin{align} \label{gpsi}
\langle \tilde{\Gamma}_i^\epsilon \vert \psi \rangle &= \left[ \frac{1}{1 - \langle \tilde{\Gamma^\epsilon} \vert (H_0 - E)^{-1} \vert \tilde{\Gamma^\epsilon} \rangle} \right]_{ij} \langle \tilde{\Gamma}_j^\epsilon \vert (H_0 - E)^{-1} \vert \varphi \rangle \,.
\end{align}
After plugging equation~(\ref{gpsi}) into equation~(\ref{psi}), and rescaling back by $\sqrt{\lambda_i/L_i}$, the resolvent takes the following form:
\begin{align}
(H - E)^{-1} &= (H_0 - E)^{-1} + \frac{1}{\sqrt{L_i L_j}} (H_0 - E)^{-1} \vert \Gamma_i^\epsilon \rangle \Phi_{ij}^{-1} \langle \Gamma_j^\epsilon \vert (H_0 - E)^{-1} \,,  
\end{align}
where $\Phi_{ij}$ refers to the principal operator, and is given by
\begin{align} \label{phi}
\Phi_{ij} &= \left\{ \begin{array}{l} \frac{1}{\lambda_i} - \frac{1}{L_i} \langle \Gamma_i^\epsilon \vert (H_0 - E)^{-1} \vert \Gamma_i^\epsilon \rangle \,, \\ - \frac{1}{\sqrt{L_i L_j}} \langle \Gamma_i^\epsilon \vert (H_0 - E)^{-1} \vert \Gamma_j^\epsilon \rangle \,. \end{array} \right.
\end{align}
By a regularized resolvent, it is meant that there is a cutoff on the high energy modes in the principal operator. Since we are studying the model on a Riemannian manifold, it is more convenient to work with heat kernel methods rather than to utilize standard momentum space formalism for renormalization. The free resolvent can be written in terms of the heat kernel:
\begin{align} \label{h}
\langle x \vert (H_0 - E)^{-1} \vert y \rangle &= \int_0^\infty \frac{dt}{\hbar} e^{E t/ \hbar} K_t (x,y) \,,
\end{align}
where the heat kernel is the fundamental solution of the heat equation for the free Hamiltonian,
\begin{align} 
\hbar \frac{\partial}{\partial t} K_t(x,y) - \frac{\hbar^2}{2m} \nabla_g^2 K_t(x,y) &= 0 \,.
\end{align}
Now, the advantage of this formalism is nothing but to replace the momentum cutoff by a cutoff on the lower limit of the integral above due to the fact that the high energy behavior corresponds to the short-time behavior of the heat kernel. In other words, the divergence encountered in momentum space calculations matches the singular behavior of the diagonal heat kernel near $t \rightarrow 0^+$ . We  use the following initial condition of the heat kernel:
\begin{align}
\lim_{t\rightarrow 0^+} K_t(x,y) = \delta_g(x,y) \,,
\end{align}
which would make the family of Hamiltonians converge to the original problem. The heat kernel has a uniformly convergent expansion in terms of the eigenfunctions of the Laplacian on compact manifolds. In the case of compact manifolds, orthogonality, completeness and the eigenfunction expansion relations are given by
\begin{align}
\delta_{\rho \sigma} &= \int_\mathcal{M} d_g^3 x f_\rho(x) f_\sigma (x) \,, \\
\delta_g(x,y) &= \sum_{\sigma} f_\sigma(x) f_\sigma (y) \,, \\
\psi(x) &= \sum_{\sigma} f_\sigma(x) \int_\mathcal{M} d_g^3 y f_\sigma(y) \psi(y) \,.
\end{align}
$K_t(x,y)$ is then expressed as
\begin{equation}
K_t(x,y)=\sum_{\sigma} e^{-\sigma t} f_\sigma(x)f_\sigma(y) \,.
\end{equation}
Now, let us look at the case where there exists only one curve. By plugging equation~(\ref{h}) into equation~(\ref{phi}), and explicitly evaluating the resulting expression in a coordinate bases, the
regularized principal operator is given by
\begin{align}
\Phi_\epsilon(E) &= \frac{1}{\lambda(\epsilon)} - \frac{1}{L} \iint_\mathcal{M} d_g^3 x \, d_g^3 y \iint_{\Gamma \times \Gamma} d_g s \, d_g s' \nonumber \\
& \qquad \quad \times \int_0^\infty \frac{dt}{\hbar} e^{E t / \hbar} K_{\epsilon/2} \left( \gamma(s) ,x \right) K_t (x,y) K_{\epsilon/2} \left( y, \gamma(s') \right) \,. 
\end{align} 
The semigroup property of the heat kernel allows us to combine the convoluted heat kernels, and shifting the time parameter then gives
\begin{align}
\Phi_\epsilon(E) &= \frac{1}{\lambda(\epsilon)} - \frac{1}{L} \iint_{\Gamma \times \Gamma} d_g s \, d_g s' \int_\epsilon^\infty \frac{dt}{\hbar} e^{(E - \epsilon) t / \hbar} K_t \left( \gamma(s) ,\gamma(s') \right) \,.
\end{align}
Since the poles of the resolvent corresponds to bound states, and the resolvent formula contains the inverse of the principal operator, zero eigenvalues of the principal operator must determine the bound-state spectrum of the model. After the renormalization of this operator, the physical bound-state energy can be obtained through a well-defined expression. Let us choose the coupling constant in such a way that after taking the limit $\epsilon \rightarrow 0^+$, the divergence can be eliminated: 
\begin{align}
\frac{1}{\lambda(\epsilon)} &= \frac{1}{\lambda_R(\mu)} + \frac{1}{L} \iint_{\Gamma \times \Gamma} d_g s \, d_g s' \int_\epsilon^\infty \frac{dt}{\hbar} e^{(-\mu^2 - \epsilon) t / \hbar} K_t \left( \gamma(s) ,\gamma(s') \right) \,, 
\end{align}
where $\lambda_R(\mu)$ is the renormalized coupling constant and $-\mu^2$ is an energy scale, which could be eliminated in favor of the physical bound-state energy. Plugging this equality into regularized principal operator gives
\begin{align}
\Phi_\epsilon &= \frac{1}{\lambda_R(\mu)} + \frac{1}{L} \iint_{\Gamma \times \Gamma} d_g s \, d_g s' \int_\epsilon^\infty \frac{dt}{\hbar} e^{- \epsilon t / \hbar} \left[ e^{ - \mu^2 t / \hbar} - e^{E t / \hbar} \right] K_t \left( \gamma(s) ,\gamma(s') \right) \,.
\end{align}
By taking the limit $\epsilon \rightarrow 0^+$ allows us to obtain the renormalized principal operator at the energy scale $-\mu^2$, is given by
\begin{align} \label{rphi}
\Phi_R(E) &= \frac{1}{\lambda_R(\mu)} + \frac{1}{L} \iint_{\Gamma \times \Gamma} d_g s \, d_g s' \int_0^\infty \frac{dt}{\hbar} \left[ e^{ - \mu^2 t / \hbar} - e^{E t / \hbar} \right] K_t \left( \gamma(s) ,\gamma(s') \right) \,.
\end{align}
This expression is valid for $E <0$. On the other hand, for the positive values of $E$, the proper analytically continued form should be used. The physical bound state $E_b = -\nu_*^2$ corresponds to the solution of
\begin{align}
\Phi_R(E_b)&= 0 \, ,
\end{align}
if we are given an arbitrary scale $\mu$. By deciding on $E_b$ the bound state energy for a single curve,  we may eliminate the unphysical scale $\mu$. For bound-state problems, it is natural to choose $\mu$ to be the same as the bound state energy, by setting $\frac{1}{\lambda_R}=0$, thereby eliminating the arbitrariness of the scale.

The heat kernel may become singular in the limit $s \rightarrow s'$. Therefore, so as to justify our previous claim about the renormalization of the model, it must be proven that the second term in equation~(\ref{rphi}) stays finite under some reasonable assumptions, which means
\begin{align}
\frac{1}{L} \iint_{\Gamma \times \Gamma} d_g s \, d_g s' \int_0^\infty \frac{dt}{\hbar} \left[ e^{ - \mu^2 t / \hbar} - e^{E t / \hbar} \right] K_t \left( \gamma(s) ,\gamma(s') \right) < \infty \,.
\end{align}
We can use an upper bound for the heat kernel, which holds for a large class of manifolds~\cite{gri}. From now on, we assume that our manifold $\mathcal{M}$ belongs to a class that admits upper and lower bounds on certain types of heat kernels as stated in~\cite{gri}. For typical manifolds of interest, we have a Gaussian upper bound on the heat kernel; especially, when we have a noncompact manifold, we will have no contribution from the volume, and then the equation above becomes smaller than
\begin{align} \label{dg}
\frac{1}{L} \iint_{\Gamma \times \Gamma} d_g s \, d_g s' \int_0^\infty \frac{dt}{\hbar} \left[ e^{ - \mu^2 t / \hbar} - e^{E t / \hbar} \right] \frac{e^{- 2 m  d_g^2 \left( \gamma(s) ,\gamma(s') \right) / C' \hbar t } }{(C \hbar t / 2 m )^{3/2}} \,,
\end{align}
in which $d_g$ is the geodesic distance, $C$ and $C'$ are some constants, whose values are not important for the sake of renormalizability. For compact manifolds, there is a term proportional to $1/V(\mathcal{M})$ in the off-diagonal upper bounds but the contribution of this term is finite. Since the singularity occurs $s \rightarrow s'$, we would like to rewrite this expression by dividing it into the part which can generate a singularity, and the rest. If one can show that the former stays finite, then the expression around the whole curve is free of divergences. It basically means that our renormalization makes sense. Equation~(\ref{dg}) can be rewritten as
\begin{align} \label{sum}
\frac{1}{L} \int_\Gamma d_g s \left( \int_{\vert \xi \vert < \delta} d_g \xi + \int_{\vert \xi \vert > \delta} d_g \xi \right) \int_0^\infty \frac{dt}{\hbar} \left[ e^{ - \mu^2 t / \hbar} - e^{-\nu^2 t / \hbar} \right] \frac{e^{- 2 m  d_g^2 \left( \gamma(s) ,\gamma(s') \right) / C' \hbar t } }{(C \hbar t / 2 m )^{3/2}} \,,
\end{align}
in which $\xi$ is the arclength of the curve between the points $s$ and $s'$, defined as $\xi = \vert s-s' \vert$ for $\xi \in [- L/2, L/2]$. $\delta$ is the maximum value, which $\xi$ can attain in a geodesic ball, and this point will be explained later. Now, we need to find a lower bound for the metric distance so as to show that the expression can be large but finite. Since the curve under consideration is a finite-length curve, one intuitively expects that the curvature of the curve plays a role in the renormalization of the model. If one can find a comparison inequality, which involves the curvature, between the length $\xi$ of a small portion of the curve and the geodesic distance between the  corresponding points at $s$ and $s'$, then the renormalization can be accomplished by the intrinsic parameters of the model only.

Since we study the model in a Riemannian manifold, one should use the generalized version of the Frenet-Serre equations~\cite{wil}. These equations allow one to describe a curve by its arclength and generalized curvatures. As we consider three-dimensional Riemannian manifolds, our curve has only two generalized curvatures, namely its geodesic curvature and its torsion. It will be shown that its geodesic curvature is actually enough for our purpose. The Frenet-Serre equations for the curve $\gamma(s)$ are given by
\begin{align}
t^i(s) &= \frac{d \gamma^i(s)}{ds} \,, \\
\nabla_{d/ds} t^i(s) &= \kappa_g (s) n^i(s) \,, \label{rc} \\
\nabla_{d/ds} n^i(s) &= - \kappa_g (s) t^i(s) + \tau(s) b^i(s) \,, \\
\nabla_{d/ds} b^i(s) &= - \tau(s) n^i(s) \,,
\end{align}
where $t^i(s),n^i(s),b^i(s)$ are tangent, normal, and binormal unit vectors, respectively, which form the Frenet-Serre frames. Moreover, $\kappa_g(s)$ and $\tau(s)$ are the geodesic curvature and the torsion of the curve, respectively.

In the following, we will consider the geodesic polar coordinates~\cite{aubin,petersen}. In this coordinate system the metric can be written as
\begin{align}
ds^2 &= dr^2 + g_{ij}(r,\theta) d \theta^i d \theta^j \,,
\end{align}
in which the indices $i,j$ do not contain the coordinate $r$. We will restrict ourselves to the segment of the curve in question lying in a geodesic ball centered around the starting point of the segment so that the exponential map is a diffeomorphism. In order to measure how much the curve is deviated from the geodesic, which is nothing but the radial coordinate in this small neighborhood, let us look at the $r$-component of equation~(\ref{rc}), which can be given by
\begin{align}
\frac{1}{2} \frac{d}{ds} \left[ \left( t^r \right)^2 \right] &= \left( \kappa_g t^r n^r + \frac{1}{2} \frac{\partial g_{ij}}{\partial r} t^r t^i t^j \right) \,.
\end{align}
The solution takes the following form:
\begin{align} \label{tr}
t^r &= \sqrt{1 + 2 \int^s \left(  \kappa_g t^r n^r + \frac{1}{2} \frac{\partial g_{ij}}{\partial r} t^r t^i t^j \right) d_g s'} \,. 
\end{align}
The partial derivative of $g_{ij}$ with respect to $r$ is actually equal to the Hessian of the distance function at any point, and since it has a single global minimum in our convex neighborhood, this term is positive definite in a sufficiently small neighborhood of the center of the ball~\cite{petersen}. Furthermore, this neighborhood is chosen to keep $d r / ds$ positive. Therefore, the term with the derivative in equation~(\ref{tr}) is positive, whereas the term multiplied by the geodesic curvature is negative at worse. If one drops the derivative term, replaces the geodesic curvature by the maximum value  $\kappa_g^*$, which it can obtain on the curve, and replaces $t^r n^r$ by its possible maximum, $-1/2$, then the following inequality is obtained:
\begin{align}
\int^r \frac{d r'}{\sqrt{1 - \kappa_g^* s(r')}} &> s(r) \,. 
\end{align}
Let us replace $s(r')$ by the maximum value, which it can take in the ball in order to make the left-hand side bigger; then the inequality becomes
\begin{align}
\frac{r}{\sqrt{1- \kappa_g^* \delta}} &> s(r) \,.
\end{align}
We can now use this inequality so as to relate the geodesic distance between the points $s$ and $s'$ with the corresponding arclength $\xi$. The inequality we need to use is thens given by
\begin{align}
(\sqrt{1-\kappa_g^* \delta}) \xi &< d_{g} \left( \gamma(s) ,\gamma(s') \right) < \xi \text{ with }0 < \delta < 1 / 2 \kappa_g^* \,. 
\end{align}
For the first term in equation~(\ref{sum}), we have the following:
\begin{align}
\frac{2}{L} \int_\Gamma d_g s \int_0^{\delta} d_g \xi \int_0^\infty \frac{dt}{\hbar} \left[ e^{ - \mu^2 t / \hbar} - e^{E t / \hbar} \right] \frac{e^{- 2 m \left(1-\kappa_g^* \delta \right) \xi^2 / C' \hbar t } }{(C \hbar t / 2 m )^{3/2}} \,.
\end{align}
However, we need another condition on the curve that prevents the curve getting too close to itself after leaving this neighborhood. It can be stated as
\begin{align}
d_g \left( \gamma(s),\gamma(s') \right) > \Delta \text{ if } \xi > \delta \,.
\end{align}
Here, $\Delta$ can be smaller than $\delta$. This is actually true, if we make the embedding more precise: we assume that the finite-length curve is an embedding of $S^1$ as a closed Riemannian submanifold of the ambient space. Then, the well-known tubular neighborhood theorem in~\cite{spi} for compact submanifolds implies that there is a disk-bundle of sufficiently small diameter around $S^1$ and a tubular neighborhood of  the curve in the ambient manifold, such that the disk-bundle of $S^1$ is isomorphic to the tubular neighborhood of the curve. After evaluating the $t$-integral, one can, hence, obtain
\begin{align}
\frac{4m}{\hbar^2} \frac{1}{L} \sqrt{\frac{\pi C'}{C^3 \left( 1 - \kappa_g^* \delta \right)}} \int_\Gamma d_g s \int_0^{\delta} d_g \xi \frac{e^{-2 \sqrt{\frac{2m}{\hbar^2 C'} \left( 1 - \kappa_g^* \delta \right)} \mu \xi} - e^{-2 \sqrt{\frac{2m}{\hbar^2 C'} \left( 1 - \kappa_g^* \delta \right)} \sqrt{-E} \xi}}{\xi} \,.
\end{align}
The integral over $\xi$ gives a term that is just proportional to $\log(\sqrt{-E}/\mu)$ as
\begin{align}
\frac{4m}{\hbar^2} \sqrt{\frac{\pi C'}{C^3 \left( 1 - \kappa_g^* \delta \right)}} \log \left( \frac{\sqrt{-E}}{\mu} \right)  \,,
\end{align}
and the rest can be given by a difference of exponential integral functions.
\section{Existence of a lower bound for the ground-state energy} \label{s3}
We will show that there is a lower bound of the ground-state energy $E_{gr}$ for a collection of curves. Assume that the renormalization is obtained by isolating each curve and assigning a binding energy  $-\nu_{i*}^2$ to the curve $\Gamma_i$. In order to prove this assertion under the given prescription, we will use a well-known theorem in the matrix analysis, i.e. the Ger\u{s}gorin theorem~\cite{mat}. It states that all the eigenvalues $\omega$ of a matrix $\Phi \in M_N$ are located in the union of $N$ disks:
\begin{align}
\bigcup_{i=1}^N \vert \omega - \Phi_{ii} \vert \leqslant \bigcup_{i=1}^N \sum_{i \neq j = 1}^N \vert \Phi_{ij} \vert \,.
\end{align}
If there is a lower bound on the ground-state energy, say $E^*$ in our problem, one expects that $\omega = 0$ is not the eigenvalue at all beyond this lower bound, $E \leqslant E^*$. In~\cite{et}, it is shown that the following inequality
\begin{align} \label{inge}
\vert \Phi_{ii} (E) \vert^{\min} > (N-1) \vert \Phi_{ij} (E) \vert^{\max} \,, 
\end{align}
implies that the ground-state energy should be larger than a lower bound $E^*$, which saturates the above inequality:
\begin{align}
E_{gr} \geqslant E^* \,.
\end{align}
For the left-hand side of the inequality~(\ref{inge}), one needs to use a lower bound for the heat kernel~\cite{dav} (for simplicity we assume manifolds with nonnegative Ricci curvature) so as to obtain a lower bound for the absolute value of the on-diagonal part of the principal operator:
\begin{align}
\vert \Phi_{ii} (E) \vert > \frac{1}{L_i} \iint_{\Gamma_i \times \Gamma_i} d_g s \, d_g s' \int_0^\infty \frac{dt}{\hbar} \left[ e^{ - \nu_{i*}^2 t / \hbar} - e^{E t / \hbar} \right] \frac{e^{- 2 m  d_g^2 \left( \gamma_i(s) ,\gamma_i(s') \right) / D' \hbar t } }{(D \hbar t / 2 m )^{3/2}} \,,  
\end{align}
where $D,D'$ are some constants. We note that a geodesic is length minimizing, i.e. the length of a geodesic is less than or equal to any admissible curve with the same endpoints. Therefore, the geodesic distance between the points $\gamma(s)$ and $\gamma(s')$ must satisfy the following  inequality:
\begin{align}
d_g \left( \gamma(s) , \gamma(s') \right) < \xi \,.
\end{align}
The on-diagonal part of the principal operator then obeys
\begin{align} \label{pii}
\vert \Phi_{ii} (E) \vert &> \frac{2}{L_i} \int_{\Gamma_i} d_g s \int_0^{L_i/2} d_g \xi \int_0^\infty \frac{dt}{\hbar} \left[ e^{ - \nu_{i*}^2 t / \hbar} - e^{E t / \hbar} \right] \frac{e^{- 2 m \xi^2 / D' \hbar t } }{(D \hbar t / 2 m )^{3/2}} \nonumber \\
\qquad &= \frac{4m}{\hbar^2} \frac{1}{L_i} \sqrt{\frac{\pi D'}{D^3}} \int_{\Gamma_i} d_g s \int_0^{L_i/2} d_g \xi \frac{e^{-2 \sqrt{\frac{2m}{\hbar^2 D'}} \nu_{i*} \xi} - e^{-2 \sqrt{\frac{2m}{\hbar^2 D'}} \sqrt{- E} \xi}}{\xi} \,.
\end{align}
One can rewrite the $\xi$-integral as $\int_0^\infty d \xi - \int_{L_i/2}^\infty d \xi$. Since we would like to minimize this expression as long as it stays positive, we need to make the second integral bigger.  By a change of variables, the lower limit of the second integral becomes $0$, and doing some manipulations afterwards, the second integral is given by
\begin{align}
& \int_0^\infty d_g\xi \frac{e^{-2 \sqrt{\frac{2m}{\hbar^2 D'}} \nu_{i*} \left(\xi + \frac{L_i}{2} \right)} - e^{- \sqrt{\frac{2m L_i^2}{\hbar^2 D'}} \nu_{i*}} e^{-2 \sqrt{\frac{2m}{\hbar^2 D'}} \sqrt{- E} \xi}}{\xi + \frac{L_i}{2}} \nonumber \\ 
& \qquad \qquad \qquad + \int_0^\infty d_g \xi \frac{e^{-2 \sqrt{\frac{2m}{\hbar^2 D'}} \sqrt{- E} \xi}}{\xi+\frac{L_i}{2}} \left[ e^{- \sqrt{\frac{2m L_i^2}{\hbar^2 D'}} \nu_{i*} } - e^{-\sqrt{\frac{2m L_i^2}{\hbar^2 D'}} \sqrt{- E}}\right] \,.
\end{align}
Dropping both the second term and $\xi$ in the denominator in the second integral makes this expression bigger. Exponentiating of the denominator in the first integral by an integral representation and performing a partial integration successively leads to
\begin{align}
& e^{- \sqrt{\frac{2m L_i^2}{\hbar^2 D'}} \nu_{i*}} \left[ \log \left( \frac{\sqrt{-E}}{\nu_{i*}} \right)  + \sqrt{\frac{\hbar^2 D'}{2 m L_i^2}} \frac{1}{\sqrt{-E}} \right] \nonumber \\ 
& \qquad \qquad \qquad - \frac{L_i}{2} e^{- \sqrt{\frac{2m L_i^2}{\hbar^2 D'}} \nu_{i*}} \int_0^\infty d \tau e^{- \tau \frac{L_i}{2}} \log \left( \frac{\tau + 2 \sqrt{\frac{2 m}{\hbar^2 D'}} \sqrt{-E}}{\tau + 2 \sqrt{\frac{2 m}{\hbar^2 D'}} \nu_{i*}} \right) \,.
\end{align} 
Since $\sqrt{-E}$ is larger than $\nu_{i*}$, the logarithm inside the integral is positive. Thereof, we obtain a bigger quantity when we drop this integral. As nothing now depends on the parameter $s$ in equation~(\ref{pii}), the $s$-integral can easily be taken, and the result is 
\begin{align}
\vert \Phi_{ii} (E) \vert &> \frac{4m}{\hbar^2} \sqrt{\frac{\pi D'}{D^3}} \left\{ \left[ 1 - e^{- \sqrt{\frac{2m L_i^2}{\hbar^2 D'}} \nu_{i*}} \right] \log \left(\frac{\sqrt{-E} }{\nu_{i*} }\right) - \sqrt{\frac{\hbar^2 D'}{2 m L_i^2}} \frac{1}{\sqrt{-E}} e^{- \sqrt{\frac{2m L_i^2}{\hbar^2 D'}} \nu_{i*}} \right\} \,.
\end{align}
The minimum value of the left-hand side of the inequality~(\ref{inge}) can, therefore, be attained as follows:
\begin{align}
\vert \Phi_{ii} (E) \vert^{\min} &> \frac{4m}{\hbar^2} \sqrt{\frac{\pi D'}{D^3}} \left\{ \left[ 1 - e^{-\sqrt{\frac{2m L_{\min}^2}{\hbar^2 D'}} \nu_*^{\min}} \right] \log \left(\frac{\sqrt{-E} }{\nu_*^{\max} }\right) - \sqrt{\frac{\hbar^2 D'}{2 m L_{\min}^2}} \frac{1}{\sqrt{-E}} e^{-\sqrt{\frac{2m L_{\min}^2}{\hbar^2 D'}} \nu_*^{\min}} \right\} \,,
\end{align}
where $L_{\min}=\min_i (L_i)$, $\nu_*^{\min}=\min_i  (\nu_{i*})$, and $\nu_*^{\max}=\max_i  (\nu_{i*})$. 

For the right-hand side of the inequality~(\ref{inge}), one needs to use an upper bound for the heat kernel for noncompact manifolds as in equation~(\ref{dg}):
\begin{align}
\vert \Phi_{ij} (E) \vert &< \frac{1}{\sqrt{L_i L_j}} \iint_{\Gamma_i \times \Gamma_j} d_g s \, d_g s' \int_0^\infty \frac{dt}{\hbar} e^{E t / \hbar} \frac{e^{- 2 m  d_g^2 \left( \gamma_i(s) ,\gamma_j(s') \right) / C' \hbar t } }{(C \hbar t / 2 m )^{3/2}} \,.
\end{align}
If one defines a minimum distance between each curve by
\begin{align}
d_{ij} &= d ( \Gamma_i, \Gamma_j) = \min_{s \in \Gamma_i,s' \in \Gamma_j} d_g ( \gamma_i(s), \gamma_j(s') ) \,,
\end{align}
then the right-hand side of the inequality can be made bigger by this distance as
\begin{align}
\vert \Phi_{ij} (E) \vert &< \frac{1}{\sqrt{L_i L_j}} \iint_{\Gamma_i \times \Gamma_j} d_g s \, d_g s' \int_0^\infty \frac{dt}{\hbar} e^{E t / \hbar}\frac{e^{- 2 m  d_{ij}^2 / C' \hbar t } }{(C \hbar t / 2 m )^{3/2}} \\
&= \frac{4m}{\hbar^2} \sqrt{\frac{\pi C'}{C^3}} \frac{1}{\sqrt{L_i L_j}} \iint_{\Gamma_i \times \Gamma_j} d_g s \, d_g s' \frac{e^{-2 \sqrt{\frac{2m}{\hbar^2 C'}} \sqrt{-E} d_{ij}} }{d_{ij}} \,.
\end{align}
After taking the $s$-integrals and using obvious inequality for $L_{\max}=\max_i (L_i)$ and $d_{\min}=\min_{i \neq j} (d_{ij})$, one obtains
\begin{align}
\vert \Phi_{ij} (E) \vert^{\max} &< \frac{4m}{\hbar^2} \sqrt{\frac{\pi C'}{C^3}} L_{\max} \frac{e^{-2 \sqrt{\frac{2m}{\hbar^2 C'}} \sqrt{-E} d_{\min}} }{d_{\min}} \,.
\end{align}

Therefore, the inequality~(\ref{inge}) takes the following form:
\begin{align}
& \left[ 1 - e^{-\sqrt{\frac{2m L_{\min}^2}{\hbar^2 D'}} \nu_*^{\min}} \right] \log \left(\frac{\sqrt{-E} }{\nu_*^{\max} }\right) - \sqrt{\frac{\hbar^2 D'}{2 m L_{\min}^2}} \frac{1}{\sqrt{-E}} e^{-\sqrt{\frac{2m L_{\min}^2}{\hbar^2 D'}} \nu_*^{\min}}  \nonumber \\ & \qquad \qquad \qquad > (N-1) \sqrt{\frac{C' D^3}{D' C^3}}\frac{L_{\max}}{d_{\min}} e^{-2 \sqrt{\frac{2m d_{\min}^2}{\hbar^2 C'}} \sqrt{-E}} \,.
\end{align}
Since the left-hand side of this inequality is an increasing function of $E$, whereas the right-hand side is a decreasing function of $E$, one can always find a solution of this inequality so that it implies the existence of a lower bound of the ground-state energy $E_{gr} \geqslant E^*$. Indeed, as long as the curves are not extremely large in general, and the minimum distance between the curves is comparable with the lengths of the curves, this inequality implies that the lower bound behaves as the inverse function of $\nu \log(\nu)$. For compact manifolds, under the same nonnegativity assumption of the Ricci curvature, we will have a volume contribution to the upper bound. This contribution is an additional $1/V(\mathcal{M})$-term, and it leads to an exponential decaying contribution proportional to $1/\sqrt{-E}$ so it does not change the characteristic of the inequality. We remark that in the case of Cartan-Hadamard manifolds, i.e. manifolds with non-positive sectional curvature with sectional curvature bounded from below, the lower bound on the diagonal $\Phi_{ii}(E)$ will be modified. The change can, essentially, be expressed by replacing $\nu_i$ and $-E$ in the lower bound by effective values, $\sqrt{\nu_i^2+a^2}+k$, and similarly for $-E$, where $a$ mainly depends on the (absolute value of the) upper bound on the sectional curvature, and $k$ depends on the lower bound of the sectional curvature~\cite{grigoryan}. We can prove the finiteness of ground-state energy in this case under similar restrictions.
\section{Positivity of the ground-state} \label{s4}
For a proof of the positivity of the ground-state, we will follow~\cite{et}. Since the procedure is
similar and the calculations repeat themselves, we will just outline the steps. The reader is
invited to refer to~\cite{et} for the details. In order to show that the ground state is positive, the first step one needs to do is to obtain the flow of the eigenvalues of the principal operator. We will resort to the well-known Feynman-Hellman theorem~\cite{fh} to see how the eigenvalues of the principal operator change with the energy $E$. Let us say $\Phi_{ij}$ satisfies the following eigenvalue equation for the $k$th eigenvalue $\omega^{(k)}(E)$:
\begin{align}
\Phi_{ij}(E) A_j^{(k)} &= \omega^{(k)}(E) A_i^{(k)} \,,
\end{align}
with $A^{(k)}$ being the $k$th eigenvector, and there is a summation over the repeated index $j$. The derivative of the principal operator with respect to the energy $E$ reads
\begin{align}
\frac{\partial \Phi_{ij}(E_{gr})}{\partial E} &= - \frac{1}{\sqrt{L_i L_j}} \iint_{\Gamma_i \times \Gamma_j} d_g s \, d_g s' \int_0^\infty \frac{dt}{\hbar} \frac{t}{\hbar} e^{E_{gr} t / \hbar} K_t \left( \gamma_i(s),\gamma_j(s') \right) \,.
\end{align}
By the Feynman-Hellman theorem, the expectation value of this expression should be equal to the derivative of the eigenvalues of the operator with respect to the energy $E$, which we rewrite in a convenient form       
\begin{align} \label{fo}
\frac{\partial \omega^{(k)}(E_{gr})}{\partial E} &= - \int_\mathcal{M} d_g^3 x \int_0^\infty \frac{dt}{\hbar} \frac{t}{\hbar} e^{E_{gr} t/\hbar} \left \vert \sum_i \frac{1}{\sqrt{L_i}}  \int_{\Gamma_i} ds K_t \left( \gamma_i(s),x \right) A_i^{(k)}  \right \vert^2 \,.
\end{align}        
It is obvious that the expression above is strictly negative. Therefore, all eigenvalues are decreasing functions of energy. This indicates that the ground-state energy must correspond to the zero of the lowest eigenvalue of the principal operator.

Now, what is known is that the wavefunction can be obtained from the resolvent equation itself. Since the eigenvalues are isolated, the integral around the contour enclosing the isolated eigenvalue defines the projection operator to the subspace corresponding to this eigenvalue. Furthermore, the free resolvent does not contain any poles on the negative real axis, and these poles must come from the poles of the inverse principal operator. The spectral decomposition of the inverse principal operator,
\begin{align}
\Phi_{ij}^{-1} &= \sum_k \frac{A_i^{(k)} A_j^{(k)}}{\omega^{(k)}} \,,
\end{align}
allows one to rewrite that part of the full resolvent for the eigenvalue $\omega^0$ as
\begin{align}
\frac{1}{\sqrt{L_i L_j}} \frac{1}{H_0 - E_{gr}} \vert \Gamma_i \rangle \frac{A_i^{(0)} A_j^{(0)}}{\omega^{(0)}} \langle \Gamma_j \vert \frac{1}{H_0 - E_{gr}} \,.
\end{align}
After the residue calculation, the ground-state wave function reads
\begin{align}
\psi_{gr}(x) &= \frac{1}{\left \vert \frac{\partial \omega (E_{gr})}{\partial E} \right \vert^{1/2}} \sum_i \frac{1}{\sqrt{L_i}} \int_0^\infty \frac{dt}{\hbar} e^{E_{gr} t / \hbar}\int_{\Gamma_i} d_g s K_t \left( x, \gamma_i(s) \right) A_i^{(0)} \,.  
\end{align}
Since $|\partial \omega / \partial E|^{1/2}$ has $1/\sqrt{L_i}$, which is obvious from equation~(\ref{fo}), and the above equation has the same factor, the wavefunction depends on the ratio of the lengths of the curves. The next step is to determine the sign of the eigenvector $A^{(0)}$. The off-diagonal elements of the principal operator matrix are negative. If one subtracts the maximum entry of the diagonal part, corresponding to the ground-state energy $E_{gr}$ from each on-diagonal element, then the
total matrix can be made strictly negative. By means of the Perron-Frobenius theorem~\cite{mat}, each element of the eigenvector corresponding to the lowest eigenvalue can be chosen to be positive. Since that matrix is obtained just by subtracting an identity matrix multiplied by a number from the original principal matrix, the latter should have the same eigenvector as the former. This implies that the eigenvector $A^{(0)}$ has all components positive. If we look at the ground-state wave function, we see that each term in it is positive:
\begin{align}
\psi_{gr}(x) &= \mathcal{N} \sum_i \frac{1}{\sqrt{L_i}} \int_0^\infty \frac{dt}{\hbar} \stackrel{>0}{\overbrace{e^{E_{gr} t / \hbar}}} \int_{\Gamma_i} ds \stackrel{>0}{\overbrace{K_t \left( x, \gamma_i(s) \right)}} \stackrel{>0}{\overbrace{A_i^{(0)}}} \,.  
\end{align}
The ground state is, hence, proven to be positive, and as a result unique. 
\section{Renormalization group equation} \label{s5}
In this section, the renormalization group equation and the beta function of the model will be derived. For this purpose, we will use another renormalization prescription, which is different from the one needed for the similar analysis for the bound state. As the renormalized principal operator is basically the principal operator, but free of divergences in the limit $s \rightarrow s'$, one defines the renormalized version by subtracting the singular part of the unrenormalized operator from itself. Since our model consists of closed curves embedded in a Riemannian manifold, this suggests to analyze the singularity structure of the model from a submanifold point of view. In this approach, the curves can be treated as various embedded circles. More precisely, we consider them as Riemannian submanifolds of the ambient manifold. Therefore, one can locally choose orthonormal frame bundles over the Riemannian manifold, which carry adapted frames of the curves~\cite{wil}. Those adapted frames are decomposed into a vector that is tangent to the curves, and the others that are normal to the curves. Since the divergence only occurs while moving along the curve in the coincidence limit, there is no geometric contribution to it, which comes from the directions related to the normal bundle. This means that one can at least locally write the heat kernel as a direct product of heat kernels as long as the coincidence limit is concerned. These heat kernels are nothing but the one on a circle with a metric coming from the embedding, and the one for the normal directions. That we are only moving along the curve allows us to have the freedom to choose the heat kernel of $\mathbb{R}^2$ in the normal directions for the removal of the singularity in the expression. In light of the above discussion, the renormalized principal operator takes the following form:
\begin{align} \label{rg}
\Phi_R(E) &= \frac{1}{\lambda_R(\mu)} + \frac{1}{\int_\Gamma d_g s} \int_0^\infty \frac{dt}{\hbar} \left[ \iint_{\Gamma \times \Gamma} d_g \theta \, d_g \theta' \frac{e^{-\mu^2 t / \hbar}}{4 \pi t} K_t^{S^1} \left( \gamma(\theta), \gamma(\theta'); e^*(g) \vert_{S^1} \right) \right.  \nonumber \\
& \qquad \qquad - \left. \iint_{\Gamma \times \Gamma} d_g s \, d_g s' K_t \left( \gamma(s),\gamma(s');g \right) e^{E t/\hbar}\right] \,,
\end{align}
in which $d_g \theta = \sqrt{e^*(g) \vert_{S^1}} d \theta$, and $e^*(g) \vert_{S^1}$ is a pull-back metric. For simplicity, we work with a single curve. In order to obtain the beta function, we need to use the renormalization group equation
\begin{align}
\mu \frac{d}{d \mu} \Phi_R(E) &=0 \,,
\end{align}
which implies that the physics is independent of the chosen scale $\mu$. If the time parameter $t$ is scaled by $\mu^{-2}$, and the scaling property of the heat kernel, i.e. $K_t(x,y;g) = \tau^{-d} K_{\tau^{-2} t} (x,y; \tau^{-2} g)$, where $d$ is the dimension of the space, is used, then straightforward calculations give rise to the following equation for the beta function:
\begin{align} \label{b}
\beta(\lambda_R) &= - \frac{\lambda_R^2}{2 \pi} \frac{1}{\int_\Gamma d_{\mu^2 g} s} \iint_{\Gamma \time \Gamma} d_{\mu^2 g} \theta d_{\mu^2 g} \theta' \int_0^\infty \frac{dt}{\hbar^2} e^{-t / \hbar} K_t \left( \gamma(\theta), \gamma(\theta'); \mu^2 e^*(g)\vert_{S^1} \right) \,. 
\end{align}
That the metric is invariant under the full group of Diff $(S^1)$ allows us to redefine the coordinate $\theta$ as
\begin{align}
\tilde{\theta} &= \int_0^\theta \sqrt{e^*(g(\theta')) \vert_{S^1}} d \theta' \,,
\end{align}
which removes the metric dependence in such a way that the metric can be converted into the standard metric on $S^1$, and so the heat kernel in equation~(\ref{b}) is given by the standard heat kernel on $S^1$ \cite{ros} as 
\begin{align}
K_t ( \tilde{\theta},\tilde{\theta}') &= \sum_{n=-\infty}^\infty \frac{e^{-\frac{n^2 \pi^2 t}{L^2 \hbar}} e^{\frac{i n \pi (\tilde{\theta} - \tilde{\theta}')}{L}}}{L} \,.
\end{align}
The beta function becomes metric independent, and only the length of the curve plays a distinctive role for the beta function thereof. As a result, we can briefly rewrite the beta function as
\begin{align}
\beta(\lambda_R) &= - \frac{\lambda_R^2}{2 \pi L} C \,,
\end{align}  
where $C$ is a constant, and stands for the integral. It is obvious that the model is asymptotically free. How the change of the energy scale effects the coupling constant can be observed by integrating the beta function. Hence, the flow of the coupling constant takes the following form:
\begin{align}\label{beta}
\lambda_R(\tau \mu) &= \frac{\lambda_R(\mu)}{1 + \frac{\lambda_R(\mu)}{2 \pi L} C \log \tau} \,.
\end{align}
In the following, it will be demonstrated that the renormalized principal operator satisfies the Callan-Symanzik equation. For that purpose, it will be shown that the change of the energy scale is equivalent to the scaling of the energy and of the metric in the expression. If we scale $E$ and $g$ as
\begin{align}
E \rightarrow \tau^2 E & \quad \text{and} \quad g \rightarrow \tau^{-2} g \,,
\end{align}
the principal operator becomes
\begin{align}
\Phi_R(\tau^2 E) &= \frac{1}{\lambda_R(\mu)} + \frac{1}{\int_\Gamma d_g s \tau^{-1}} \int_0^\infty \frac{dt}{\hbar} \left[ \iint d_g \theta \, d_g \theta' \tau^{-2} \frac{e^{-\mu^2 t / \hbar}}{4 \pi t} K_t^{S^1} \left( \gamma(\theta), \gamma(\theta'); \tau^{-2}e^*(g) \vert_{S^1} \right) \right. \nonumber \\
& \qquad - \left. \iint_{\Gamma \times \Gamma} d_g s \, d_g s' \tau^{-2 }K_t \left( \gamma(s),\gamma(s');\tau^{-2} g \right) e^{\tau^2 E t/\hbar}\right] \,.
\end{align}
After scaling $t \rightarrow \tau^{-2}t$, and successively using the scaling property of the heat kernel, the principal operator is given by
\begin{align}
\Phi_R(\tau^2 E) &= \frac{1}{\lambda_R(\mu)} + \frac{1}{L} \int_0^\infty \frac{dt}{\hbar} \left[ \iint d_g \theta \, d_g \theta' \frac{e^{-\mu^2 \tau^{-2} t / \hbar}}{4 \pi t} K_t^{S^1} \left( \gamma(\theta), \gamma(\theta'); e^*(g) \vert_{S^1} \right) \right. \nonumber \\
& \qquad - \left. \iint_{\Gamma \times \Gamma} d_g s \, d_g s' K_t \left( \gamma(s),\gamma(s');g \right) e^{E t/\hbar}\right] \,. 
\end{align}
Therefore, the following result can be inferred from the equation above: 
\begin{align}
\Phi_R \left( \mu, \lambda_R(\mu), \tau^2 E, \tau^{-2} g \right) &= \Phi_R \left( \mu \tau^{-1} , \lambda_R(\mu), E, g \right) \,.
\end{align}
Taking the scale-invariant derivative with respect to $\tau$ of both sides leads to the renormalization group equation for the principal operator,
\begin{align}
\left[ \tau \frac{d}{d \tau} - \beta(\lambda_R) \frac{\partial}{\partial \lambda_R} \right] \Phi_R \left( \mu, \lambda_R(\mu), \tau^2 E, \tau^{-2} g \right) &=0 \,.
\end{align}
The following ansatz for the operator $\Phi_R(E)$:         
\begin{align}
\Phi_R \left( \mu, \lambda_R(\mu), \tau^2 E, \tau^{-2} g \right) &= f(\tau) \Phi_R \left( \mu , \lambda_R(\mu \tau), E, g \right) \,,
\end{align}         
gives that         
\begin{align}
\tau \frac{d}{d \tau} f(\tau) &=0 \,,
\end{align}
whose solution is $f(\tau)=1$ by means of the initial condition $f(\tau=1)=1$. Therefore, we obtain the desired equation for the renormalized principal operator. It suggests that the use of the renormalized coupling constant at a different energy scale is equivalent to the scaling of the energy and of the metric in this expression at the same energy scale $\mu$:
\begin{align}\label{sca}
\Phi_R \left( \mu, \lambda_R(\mu), \tau^2 E, \tau^{-2} g \right) &= \Phi_R \left( \mu , \lambda_R(\mu \tau), E, g \right) \,.
\end{align}
We can explicitly verify that if the renormalized coupling constant $\lambda_R$ changes according to equation~(\ref{beta}), the scaling law in equation~(\ref{sca}) is indeed satisfied.
\section{Comment on the two-dimensional case} \label{s6}
We would like to show that choosing the dimension of the ambient manifold $d=2$ leads to a finite theory so that it does not require a renormalization. For the bound-state energy $E_b=-\nu_*^2$, the principal operator just for a single curve becomes
\begin{align}
\frac{1}{\lambda} - \frac{1}{L} \iint_{\Gamma \times \Gamma} d_g s \, d_g s' \int_0^\infty \frac{dt}{\hbar} e^{- \nu_*^2 t / \hbar} K_t \left( \gamma(s), \gamma(s') \right) &=0 \,.
\end{align}
Now, we need to show that 
\begin{align}
\frac{1}{L} \iint_{\Gamma \times \Gamma} d_g s \, d_g s' \int_0^\infty \frac{dt}{\hbar} e^{- \nu_*^2 t / \hbar} K_t \left( \gamma(s), \gamma(s') \right) < \infty \,.
\end{align}          
If one uses the same upper bound for the heat kernel, having been used before, the following expression is obtained:
\begin{align}
\frac{4m}{\hbar^2 L C} \iint_{\Gamma \times \Gamma} d_g s \, d_g s' K_0 \left( 2 \nu_* d_g \left( \gamma(s),\gamma(s') \right) \sqrt{2m /\hbar^2 C'} \right) < \infty \,,
\end{align}
where $K_0$ is the modified Bessel function of order $0$. Taking the limit $d_g \rightarrow 0$ gives          
\begin{align}
\frac{4m}{\hbar^2 L C} \iint_{\Gamma \times \Gamma} d_g s \, d_g s' \log \left( \sqrt{\frac{\hbar^2 C'}{2m}} \frac{e^{- \gamma}}{\nu_* d_g \left( \gamma(s),\gamma(s') \right)} \right) < \infty \,,
\end{align} 
where $\gamma$ is the so-called Euler-gamma number. This expression is indeed finite. The generalization to the multi-curve case is exactly the same as before. One can see that even in this finite theory, the resolvent approach is more versatile in comparison to the differential equation approach. For example, if the curves are separated by large distances, one can apply a kind of perturbation theory, as presented in~\cite{et}. But a more striking example would be to consider a curve and a point interaction which is not on the curve. In this case, the strength of the interaction on the curve should be given, yet the point interaction requires a renormalization, which can be performed as explained in~\cite{et}. Investigating the spectral properties of this combined system is most naturally done in this approach.\section{Acknowledgment}
O. T. Turgut would like to thank P. Exner and M. Znojil, for their  kind invitation to Doppler Institute, Prague, where this work has begun, and also for stimulating discussions. This work is supported by Bo\u{g}azi\c{c}i University BAP Project 6513.

\end{document}